\documentclass[a4paper,11pt]{article}
\usepackage{amsfonts}
\usepackage{amsmath,amssymb}
\usepackage{graphicx}
\usepackage{amsmath,dsfont}
\usepackage[english]{babel}

\makeatletter \@addtoreset{equation}{section}
\def\be{\begin{equation}}
\def\ee{\end{equation}}
\def\ben{$$} \def\een{$$}
\def\ba{\begin{array}{c}}

\def\ea{\end{array}} 
\def\bea{\begin{eqnarray}}
\def\eea{\end{eqnarray}}

\def\ben{\begin{displaymath}}
\def\een{\end{displaymath}}
\def\ba{\begin{array}{c}}
\def\bal{\begin{array}{l}}
\def\ea{\end{array}}

\def\obj(#1)(#2)#3#4#5{%
  \psline[arrows={[-]}, linestyle=dashed, dash=0.0 1,dashadjust=false](#1)(#2)%
  \uput{0.4}[90](#1){#3}%
  \uput{0.4}[-90](#1){#4}\uput{0.4}[-90](#2){#5}%
}
\def\objd(#1)(#2)#3#4#5{%
  \psline[arrows={[-}, linestyle=dashed, dash=0.0 1,dashadjust=false](#1)(#2)%
  \uput{0.4}[90](#1){#3}%
  \uput{0.4}[-90](#1){#4}\uput{0.4}[-90](#2){#5}%
}

\begin{document}
\title{
Supersymmetries of the spin-$1/2$ particle in the field of
magnetic vortex, and anyons}

\author{\textsf{Francisco Correa$^a$}\textsf{$\,$, Horacio Falomir$^b$}
\textsf{$\,$, V\'{\i}t Jakubsk\'y$^c$}\textsf{$\,$and
Mikhail S. Plyushchay$^{d,e}$}
\\
{\small \textit{${}^a$ Centro de Estudios Cient\'{\i}ficos (CECS), Valdivia, Chile }}\\
{\small \textit{${}^b$ IFLP/CONICET   -   Departamento   de   F\'{\i}sica,
 Facultad de Ciencias Exactas,}}\\
 {\small \textit{Universidad Nacional de La Plata, C.C. 67, (1900)  La Plata, Argentina}}\\
{\small \textit{${}^c$ Nuclear Physics Institute, ASCR, 250 68 \v Re\v z, Czech Republic}}\\
{\small \textit{${}^{d}$ Departamento de F\'{\i}sica,
Universidad de Santiago de Chile, Casilla 307, Santiago 2,
Chile  }}\\
{\small \textit{${}^e$ Departamento de F\'{\i}sica
Te\'orica, At\'omica y \'Optica, Universidad de Valladolid,
47071, Valladolid, Spain}}\\
 \sl{\small{E-mails:  correa@cecs.cl,
falomir@fisica.unlp.edu.ar, v.jakubsky@gmail.com,
mikhail.plyushchay@usach.cl}}}
\date{}
\maketitle
\begin{abstract}
The quantum nonrelativistic spin-$1/2$ planar systems in
the presence of a perpendicular magnetic field are known to
possess the $N=2$ supersymmetry. We consider such a system
in the field of a magnetic vortex, and find that there are
just two self-adjoint extensions of the Hamiltonian that
are compatible with the standard $N=2$ supersymmetry. We
show that only in these two cases one of the subsystems
coincides with the original spinless Aharonov-Bohm model
and comes accompanied by the super-partner Hamiltonian
which allows a singular behavior of the wave functions.  We
find a family of additional, nonlocal integrals of motion
and treat them together with local supercharges in the
unifying framework of the tri-supersymmetry. The inclusion
of the dynamical conformal symmetries leads to an
infinitely generated superalgebra, that  contains several
representations of the superconformal $osp(2|2)$ symmetry.
We present the application of the results in the framework
of the two-body model of identical anyons. The nontrivial
contact interaction and the emerging $N=2$ linear and
nonlinear supersymmetries of the anyons are discussed.
\end{abstract}

\section{Introduction}

It is well known that a non-relativistic planar system of a
spin-$1/2$ particle of gyromagnetic ratio $g=2$ in the
presence of an external perpendicular magnetic field
$\vec{B}=B_3(x_1,x_2) \hat{x}_3$ possesses $N=2$
supersymmetry \cite{AhCas,GK,CKS}. Its dynamics is
described by the Pauli Hamiltonian
\begin{equation}
     \mathcal{H}=\frac{1}{2m}\sum_{j=1,2} \mathcal{P}^2_j
    -\frac{e\hbar}{2mc}B_3\, \sigma_3\,,\label{PH}
\end{equation}
where $\mathcal{P}_{j}=-i \hbar
\partial_j-\frac{e}{c}A_{j}$, $B_3=\partial_1
A_2-\partial_2A_1$, that can be presented as a perfect
square~\footnote{We set $m=1/2$, $\hbar=c=-e=1$ from now
on.}
\begin{equation}\label{1-2}
     \mathcal{H}=Q_1^2\,,\qquad
     Q_1=\sum_{j=1,2} \mathcal{P}_j
     \sigma_j \,.
\end{equation}
The $Q_1$ is therefore an integral of motion. The spin
$s_3=\frac{1}{2}\sigma_3$  is another conserved quantity.
Due to relations $\sigma_3^2=1$ and $\{\sigma_3,Q_1\}=0$,
the $\Gamma=\sigma_3$ can be considered as the
$\mathbb{Z}_2$-grading operator, while $Q_1$ can be treated
as a fermionic supercharge. Another supercharge is then
defined as
\begin{equation}\label{Q2sigma}
     Q_2=i\sigma_3Q_1=\epsilon_{ij}\mathcal{P}_i\sigma_j
     =\mathcal{P}_1\sigma_2-
     \mathcal{P}_2\sigma_1\,.
\end{equation}
The $Q_a$, $a=1,2$,  and $\mathcal{H}$ generate the $N=2$
superalgebra graded by $\sigma_3$,
\begin{equation}\label{1-6}
     \{Q_a,Q_b\}=2\delta_{ab}\mathcal{H}\,,\qquad
     [Q_a,H]=0\,,\qquad
     \{\sigma_3,Q_a\}=[\sigma_3,
     \mathcal{H}]=0\,.
\end{equation}

The described algebraic procedure applies to the case of
any external field $B_3(x_1,x_2)$. For the simplest case of
a \textit{homogeneous} magnetic field, the supersymmetry
provides with a natural explanation of the degeneracy of
the Landau levels \cite{Landeg}, and it is essential in the
understanding of the quantum Hall effect \cite{QHE}.

The procedure outlined in (\ref{1-2})--(\ref{1-6}) is quite
formal, however, and can miss some important subtleties
related to the domains of the involved supersymmetry
generators.

In the present paper we focus on investigation of the
supersymmetry and its conformal extension for a
non-relativistic electron in the presence of a
\textit{magnetic vortex}.  This corresponds to the
Aharonov-Bohm (AB) effect for the spin-$1/2$
particle system, in which the indicated subtleties play a
crucial role.

The field in the case of interest is produced by an
infinitely thin solenoid of infinite length that punctures
the plane at $x_1=x_2=0$. The electromagnetic potential in
the symmetric gauge reads
\begin{equation}\label{3}
    \vec{A}=\frac{\Phi}{2 \pi}
    \left(-\frac{x_2}{x_1^2+x_2^2},
    \frac{x_1}{x_1^2+x_2^2},0\right)=
    \frac{\Phi}{2 \pi r}\left(-\sin\varphi \, ,\
    \cos \varphi \,,0\right)\,,
\end{equation}
where $x_1=r\cos\varphi, \ x_2=r\sin\varphi$,
$-\pi<\varphi\leq\pi$, and $\Phi$ is the flux of the
singular magnetic field,
 $B_3=\Phi\,\delta^2(x_1,x_2)$.
Corresponding Hamiltonian is
\begin{equation}\label{H}
  \mathcal{H}_\alpha=
            -\partial^2_r -\frac{1}{r}\partial_r+\frac{1}{r^2}(-i
            \partial_\varphi+\alpha)^2 +\alpha
            \frac{1}{r}\delta(r)\sigma_3\,,\quad \alpha=\frac{1}{2\pi}\Phi\,,
\end{equation}
where we use the identity $\delta^2(x_1,x_2)=\frac{1}{\pi
r}\delta(r)$ for the two dimensional Dirac delta
function.\vskip0.05cm

The (spinless) model was introduced by Aharonov and Bohm in
a seminal work \cite{aharonov}, where they demonstrated the
significance of the electromagnetic potential in quantum
mechanics. It is this setting that was used in the
dynamical realization of anyons \cite{wilczek},
\cite{anyonsbooks}. The relativistic modification of the
system appears in the study of the cosmic strings
\cite{cosmicstrings}, \cite{SGer}, the topological defects
which are supposed to appear in the early universe.
Nowadays, the AB effect attracts much attention in the
physics of graphene and nanotubes \cite{nanotubes}.
\vskip0.1cm

In two dimensions, the Dirac delta term in (\ref{H}) is
not, however, defined uniquely \cite{Albeverio}. The system
can be specified unambiguously as soon as the domain of the
Hamiltonian (\ref{H}) is fixed \cite{Stovicek}
\footnote{Actually, taking into account the domain of the
self-adjoint extension of the Hamiltonian, the term with
the Dirac delta can be neglected in (\ref{H}).}. Different
choices of the domain lead then to different physical
systems. Distinct physical properties are usually
attributed to the variation of the hidden characteristics
of the magnetic flux within the vortex \cite{moroz}. Here,
we identify and analyze the systems described by (\ref{H}),
for which the procedure (\ref{1-2})--(\ref{1-6}) is
consistent.\vskip0.1cm

The paper is organized as follows. In the next Section, we
find the proper domain of the operators $Q_1$ and
$Q_2=i\sigma_3Q_1$ in order to identify which kind of AB
Hamiltonians allows a physical supersymmetric structure. We
will see that the physically acceptable Hamiltonians with
supersymmetry obligatorily involve non trivial self-adjoint
extensions. So far this crucial point was not discussed
appropriately in the broad literature on the  AB problem
and supersymmetry \cite{barut}, \cite{Park},
\cite{Horvathy}. In Section 3, we identify additional
 nonlocal integrals of motion,  and analyze the
associated hidden supersymmetry and related extended
supersymmetric structure in the form of
\emph{tri-supersymmetry} \cite{finitegap}, \cite{AdS2}. The
incorporation of the  conformal symmetry into the framework
of the tri-supersymmetry is considered in Section 4. We
apply the results to the theory of anyons in Section 5.
There, the two-body systems of the interacting anyons are
discussed in the light of the superalgebraic structure
revealed in the preceding sections. In particular, the
$N=2$ nonlinear supersymmetry of the two-body anyonic model
is presented. The last Section is devoted to the discussion
of the results.

\section{$N=2$ supersymmetry}
In the forthcoming analysis of the system, the magnetic
flux will be restricted to the interval $\alpha\in[0,1)$.
This can be done without loss of generality as the systems
with the fluxes $\alpha$ and  $\alpha+n$ for integer $n$
can be related by the unitary transformation,
\begin{equation}\label{uni}
\mathcal{H}_{\alpha+n}=U_{n}^{-1}
\mathcal{H}_{\alpha}U_{n},\quad
U_{n}=e^{in\varphi}\,\mathds{1}\,,
\end{equation}
where $\mathds{1}$ is the  unit $2\times 2$ matrix.

The \emph{formal} supercharge operator $Q_1$ of the
Aharonov-Bohm system can be obtained directly by the
substitution of (\ref{3}) into (\ref{1-2}).
The explicit form of
the $Q_1$ in polar coordinates,
\begin{equation}\label{4}
    Q_1=q_+\sigma_+
            +q_-\sigma_-\,,\qquad
            q_{\pm}=-ie^{\mp i\varphi}\left(\partial_r\pm
      \frac{1}{r}\left(-i\partial_{\varphi}+\alpha\right)\right),\quad
            \sigma_\pm=\frac{1}{2}(\sigma_1\pm
            i\sigma_2)\,,
\end{equation}
suggests that it preserves the subspaces $\mathfrak{H}_l$,
\begin{equation}\label{subspace1}
    \mathfrak{H}_l=\left(\begin{array}{c} e^{il\varphi}
    \\e^{i(l+1)\varphi}
    \end{array}\right)\otimes L_2(\mathbb{R}_+,rdr),
    \qquad Q_1\mathfrak{H}_l\subset \mathfrak{H}_l,
    \qquad l\in\mathbb{Z}\, .
\end{equation}
The eigenvectors of the $Q_1$ are the states of fixed
energy, see (\ref{1-2}).
Define the function $\Phi_{\lambda,l}(r,\varphi)=(e^{il\varphi}
\phi_{\lambda,1}(r),e^{i(l+1)\varphi}\phi_{\lambda,2}(r))^T$,
where $T$ means a transposition. The eigenvalue equation
$Q_1\Phi_{\lambda,l}(r,\varphi)
=\lambda\Phi_{\lambda,l}(r,\varphi)$ can be written
equivalently as
\begin{equation}\label{10}
       \left[ \partial_r+(l+\alpha+1)\frac{1}{r}\right]
       \phi_{\lambda,2}(r)= i\lambda \phi_{\lambda,1}(r)\,, 
       \qquad
         \left[ \partial_r-(l+\alpha)\frac{1}{r}\right]
         \phi_{\lambda,1}(r)= i\lambda \phi_{\lambda,2}(r)\, .
\end{equation}
The general solution of (\ref{10}) is given in terms of the
Bessel functions of the first and of the second kinds.
Consider the case of nonzero $\lambda$. The lower component
of the eigenfunction $\Phi_{\lambda,l}(r,\varphi)$ is
\begin{equation}\label{12}
    \phi_{\lambda,2}(r)=C_1^l \mathcal{J}_{|l+\alpha+1|}
    (|\lambda| r) + C_2^l \mathcal{Y}_{|l+\alpha+1|}(|\lambda| r) \,.
\end{equation}
The associated upper component depends on $\epsilon=sign \,
(l+\alpha+1)$,
\begin{equation}\label{14}
    \phi_{\lambda,1}(r)=-i\epsilon \frac{\lambda}{|\lambda|}
    \left( C_1^l \mathcal{J}_{\epsilon(l+\alpha)}(|\lambda| r) +
    C_2^l \mathcal{Y}_{\epsilon(l+\alpha)}(|\lambda| r)\right).
\end{equation}
The solutions are not square integrable. They are
acceptable, however, in the sense of the scattering states
(generalized wave functions) as long as their \emph{square
integrability at the origin} is maintained.

For $\alpha=0$, this condition fixes $C_2^l=0$ for all
values of $l$. The wave functions $\Phi_{\lambda,l}$ are
regular at the origin and the system can be identified with
the free spin-$\frac{1}{2}$ particle in the plane. The
domain of $\mathcal{H}_{0}$ is invariant with respect to
$\sigma_3$, and the procedure (\ref{1-2})--(\ref{1-6}) is
consistent for the free-particle system.

We consider the case $\alpha\in (0,1)$ from now on if not
stated otherwise. The condition of \emph{local} square
integrability fixes $C_2^l=0$ again as long as $l\neq -1$.
However, for $l=-1$, the function $\Phi_{\lambda,-1}$ is
square integrable for any values of the coefficients
$C_1^{l}$ and $C_2^{l}$ with $l=-1$. This \textit{spoils}
its physical interpretation as the function cannot be fixed
(up to an overall normalization) uniquely. To avoid this
ambiguity, the behavior of the wave functions for $r\sim 0$
has to be specified. This, in turn, will fix the
self-adjoint extension of the $Q_1$.

The operator $Q_1$ defined  on the space of infinitely
smooth two-component functions with compact support is
symmetric. Using a standard theory of self-adjoint
extensions of symmetric operators \cite{R-S}, we extend its
domain by relaxing the regularity of the functions in the
sector of two specific partial waves. The corresponding
boundary conditions
\begin{equation}\label{boundary}
     \lim_{r\rightarrow 0^+}\Psi\sim\left(\begin{array}{l}
    (1+e^{i\gamma})2^{-\alpha}\Gamma(1 -
    \alpha) r^{-1+\alpha}e^{-i\varphi}\\
    (1-e^{i\gamma})2^{-1+\alpha}\Gamma(\alpha)
     r^{-\alpha}
    \end{array}
    \right)
\end{equation}
specify the domain of the self-adjoint extensions
$Q^{\gamma}_1$ of the $Q_1$ for $\alpha\in(0,1)$ that form
a one-parameter family marked by $\gamma\in[0,2\pi)$.

Defining the Hamiltonian
$\mathcal{H}^{\gamma}_{\alpha}=(Q_1^{\gamma})^2$, the
operator $Q_1^{\gamma}$ satisfies the relations $
[Q_1^{\gamma},\mathcal{H}^{\gamma}_{\alpha}]=0,\
\{Q_1^{\gamma},Q_1^{\gamma}\}
=2\mathcal{H}_{\alpha}^{\gamma}$,  and might be considered
as a supercharge of the $N=1$ supersymmetry. The action of
the operator $\sigma_3$ is not well defined, however, for
generic values of $\gamma$; it preserves the boundary
conditions (\ref{boundary}) if and only if the parameter
$\gamma$ acquires only two discrete values
\begin{equation}\label{gamma}
    \gamma=0,\,\pi\,.
\end{equation}
For other values of $\gamma$, the $\sigma_3$ perturbs the
relative coefficient between the up and down components of
(\ref{boundary}) and maps the wave functions out of the
physical domain. In the two cases (\ref{gamma}), the spin
operator  $s_3=\frac{1}{2}\sigma_3$ is a well defined
physical observable that is a symmetry of the Hamiltonians
\begin{equation}\label{h0}
\mathcal{H}_\alpha^{\gamma=0}=\left(
          \begin{array}{cc}
          H_\alpha^0 & 0 \\ \\
           0 & H_\alpha^{AB} \\
          \end{array}
        \right)\,,\qquad \qquad
    \mathcal{H}_\alpha^{\gamma=\pi}=\left(
          \begin{array}{cc}
          H_\alpha^{AB} & 0 \\ \\
           0 & H_\alpha^{\pi} \\
          \end{array}
        \right)\,,
\end{equation}
$[s_3,\mathcal{H}_\alpha^\gamma]=0$. These two Hamiltonians
possess therefore the $N=2$ supersymmetry since the second
supercharge $Q_2^{\gamma}= i\sigma_3Q_1^{\gamma}$ is
defined consistently on the same domain as $Q_1^{\gamma}$.
In accordance with (\ref{1-2}), (\ref{Q2sigma}), the
explicit matrix form of the supercharges is
\begin{equation}\label{Q12gamma}
     Q_1^\gamma=
     \left(\begin{array}{cc}
     0&{\cal P}_1-i{\cal P}_2\\
     {\cal P}_1+i{\cal P}_2&0\end{array}\right)\,,\qquad
     Q_2^\gamma=
     \left(\begin{array}{cc}
     0&{\cal P}_2+i{\cal P}_1\\
     {\cal P}_2-i{\cal P}_1&0
     \end{array}\right).
\end{equation}
Here, as in (\ref{h0}), for $\gamma=0$ and $\gamma=\pi$ the
corresponding operators act on different domains.

The boundary conditions (\ref{boundary}) acquire
particularly simple form\,: for $\gamma=\pi$ the upper
component disappears, while for $\gamma=0$ the lower one
does. This enables a direct interpretation of the subsystem
represented by $H_{\alpha}^{AB}$; it's domain is free of
the singular wave functions since the corresponding
singular component in the boundary condition
(\ref{boundary}) vanishes for the specific choice of
$\gamma$. In what follows, the Hamiltonian
$H_{\alpha}^{AB}$ can therefore be identified with that of
the spinless system studied originally by Aharonov and Bohm
\cite{aharonov}. We shall stress that this subsystem
appears only for (\ref{gamma}). For other values of
$\gamma$, the singular behavior is enforced in both, up-
and down-, components of the wave functions, see
(\ref{boundary}).

The generator of rotations in the plane, the orbital
angular momentum shifted by the flux,
\begin{equation}\label{orbJ}
    J=x_1\mathcal{P}_2-x_2 \mathcal{P}_1=
    -i\partial_{\varphi}+\alpha\,,
\end{equation}
represents another symmetry since it commutes with
(\ref{h0})  and preserves the boundary conditions
(\ref{boundary}) for $\gamma=0,\pi$. Hence, both systems
(\ref{h0}) are rotationally invariant. Moreover, each
subsystem $H^{AB}_\alpha$, $H_\alpha^0$ and $H^\pi_\alpha$
is rotationally invariant since $s_3$ is also the integrals
of motion. Notice here that for the values of $\gamma$
different from (\ref{gamma}), the operator $J$ would change
the relative coefficient of the up- and down- components in
(\ref{boundary}), and would not preserve the domain of
$Q_{1}^{\gamma}$. This should be interpreted as the absence
of the rotational invariance for diagonal subsystems of the
system $\mathcal{H}_\alpha^\gamma=(Q_1^\gamma)^2$,
$\gamma\neq 0,\pi$,  caused by the inner characteristics of
the magnetic flux. The total angular momentum $J+s_3$ would
preserve, however,  the boundary conditions
(\ref{boundary}) even for $\gamma\neq 0,\pi$, and would be
an integral of motion of the $N=1$ supersymmetric family of
the systems in which we are not interested anymore.

From now on, we will suppose that $\gamma$ acquires one of
the two values specified in (\ref{gamma}). In this case,
both  systems (\ref{h0}) are rotationally invariant, the
spin is preserved, and they possess the $N=2$
supersymmetry. The subsystem $H^{AB}_{\alpha}$ coincides
with the original Aharonov-Bohm setting.\vskip0.05cm

We can find the common eigenvectors $\Psi_{E,l,s}^{\gamma}$
of $\mathcal{H}^{\gamma}_{\alpha}$, $s_3$ and $J$ marked by
the energy $E$, the orbital angular momentum $l$ and
the spin sign $s=\pm $,
\begin{equation}
    \mathcal{H}_\alpha^\gamma \Psi_{E,l,s}^{\gamma}=
    E\Psi_{E,l,s}^{\gamma},\qquad
    J\Psi_{E,l,s}^{\gamma}=(l+\alpha)\Psi_{E,l,s}^{\gamma},
    \qquad
    s_3\Psi_{E,l,\pm}^{\gamma}=\pm\frac{1}{2}\Psi_{E,l,\pm}^{\gamma}\,.
\end{equation}
Having in mind the relation
$(Q_1^{\gamma})^2=\mathcal{H}_\alpha^{\gamma}$, these
functions are obtained as linear combinations of
$\Phi_{\lambda,l}$ and $\Phi_{-\lambda,l}$. The wave
functions $\Psi_{E,l,s}^{\gamma}$ for $l\neq-1$ can be
written as $\Psi_{E,l,+}^{\gamma}=\mathcal{J}_{|\alpha+l|}
(\sqrt{E}r)e^{il\varphi}\,\mathbf{v}_{+}$ and
$\Psi_{E,l+1,-}^{\gamma}=\mathcal{J}_{|\alpha+l+1|}
(\sqrt{E}r)e^{i(l+1)\varphi}\,\mathbf{v}_{-}$ where
$s_3\,\mathbf{v}_{\pm}=\pm\frac{1}{2}\,\mathbf{v}_{\pm}$.
The partial waves in the subspace $\mathfrak{H}_{-1}$ have
to obey the boundary conditions (\ref{boundary}), which fix
the coefficients in (\ref{12}) and (\ref{14}) in a unique
way. Their explicit form is
\begin{equation}\label{2}
    \Psi_{E,-1,+}^{\pi}=\left(\mathcal{J}_{1-
    \alpha}(\sqrt{E}\,r)e^{-i\varphi},
    0\right)^T,\qquad \Psi_{E,0,-}^{\pi}=\left(0
    ,\,\mathcal{J}_{-\alpha}(\sqrt{E}\,r)
    \right)^T, \qquad \gamma=\pi ,
\end{equation}
and
\begin{equation}\label{3a}
    \Psi_{E,-1,+}^{0}=\left(\mathcal{J}_{-1+\alpha}
    (\sqrt{E}\,r)e^{-i\varphi},0\right)^T,\qquad \Psi_{E,0,-}^{0}
    =\left(0,\,\mathcal{J}_{\alpha}(\sqrt{E}\,r)\right)^T , \qquad \gamma=0\,.
\end{equation}

To complete the analysis of the spectrum of
$\mathcal{H}_{\alpha}^{\gamma}$, we present the zero energy
wave functions. The partial waves with $l\neq -1$ are
$\Psi^{\gamma}_{0,l,+}=r^{|l+\alpha|} e^{il\varphi}\,
\mathbf{v}_{+}$ and
$\Psi^{\gamma}_{0,l+1,-}=r^{|l+1+\alpha|}
e^{i(l+1)\varphi}\, \mathbf{v}_{-}$ while the wave
functions of zero energy from the subspace
$\mathfrak{H}_{-1}$ acquire the form
\begin{equation}\label{E01}
    \Psi_{0,-1,+}^{\gamma}=\left\{\begin{array}{lc}(r^{1-
    \alpha}e^{-i\varphi},0)^T\,,&\gamma=\pi\\
    (r^{-1+\alpha}e^{-i\varphi},0)^T\,,&
    \gamma=0 \end{array}\right.,\qquad
    \Psi_{0,0,-}^{\gamma}=\left\{\begin{array}{lc}(0,
    r^{-\alpha})^T,&\gamma=\pi\\
    (0,r^{\alpha})^T,& \gamma=0 \end{array}\right., \quad
    E=0\,.
\end{equation}
Finally, let us note that the wave functions $\Psi_{E,l,\pm}$ for \textit{all} integer $l$
and non-negative $E$ form the basis of the Hilbert space.

Let us discuss now the action of the supercharges on the
eigenstates $\Psi_{E,l,\pm}^{\gamma}$. By the action of the
$Q_1^{\gamma}$, the spin-up and spin-down eigenstates of
positive energy are interchanged and the angular momentum
is altered,
\begin{equation}\label{action1}
     Q_{1}^{\gamma}\Psi_{E,l,+}^{\gamma}\sim \Psi_{E,l+1,-}^{\gamma}\,,
     \qquad Q_{1}^{\gamma}\Psi_{E,l+1,-}^{\gamma}\sim \Psi_{E,l,+}^{\gamma}\,.
\end{equation}
In this manner, the supercharge $Q_1^{\gamma}$ reflects the
degeneracy within the subspaces $\mathfrak{H}_l$. Each of
these subspaces contains two zero modes, $\Psi_{0,l,+}$ and
$\Psi_{0,l+1,-}$,  of the
$\mathcal{H}_{\alpha}^{\gamma}$\,; however, only one of
them is annihilated by $Q_1^{\gamma}$. \vskip0.1cm

In conclusion of this Section, let us summarize the
results. For non-integer values of $\alpha$, there are just
\emph{two} systems described either by
$\mathcal{H}_{\alpha}^{0}$ or $\mathcal{H}^{\pi}_{\alpha}$,
which possess the genuine $N=2$ supersymmetry. Only in
these two cases, one of the subsystems, represented by
$H^{AB}_{\alpha}$, coincides with the original spinless
Aharonov-Bohm model and comes accompanied with the
super-partner Hamiltonian which allows the singular
behavior of the wave functions. When $\alpha$ acquires
integer values, the studied model is unitary equivalent to
the free electron in the plane and has the $N=2$
supersymmetry as well. To our best knowledge, the careful
treatment of this kind seems to be missing in the
literature.\vskip0.1cm

In the next Section we will show that the described $N=2$
supersymmetry is just a piece of a remarkable mosaic of the
rich supersymmetric structure that underlines the system.

\section{Hidden supersymmetry and tri-supersymmetry}

The diagonal components of $\mathcal{H}^0_{\alpha}$ and
$\mathcal{H}^{\pi}_{\alpha}$, identified with one of
$H^{AB}_{\alpha}$, $H^{0}_{\alpha}$ or $H^{\pi}_{\alpha}$,
can be looked at as those describing the spinless systems
in presence of the magnetic vortex. These spinless
Hamiltonians possess $N=2$ hidden (bosonized) supersymmetry
in terms of self-adjoint, nonlocal supercharges, see
\cite{Correa}~\footnote{For earlier observations of the
hidden supersymmetry in different quantum mechanical
systems see \cite{hidsusy}, \cite{hidsusynon}.}. This
observation makes it possible to define directly two new
operators $\tilde{Q}_1$ and $\tilde{Q}_2$,
\begin{equation}\label{hiddenSUSY}
     \tilde{Q}^{\gamma}_1=\left(\begin{array}{cc}
    \mathcal{P}_1+ie^{i\gamma}R\mathcal{P}_2&0\\0&\mathcal{P}_1-
    ie^{i\gamma}R\mathcal{P}_2
     \end{array}\right)\,,\qquad
     \quad \tilde{Q}^{\gamma}_2
    =iR\tilde{Q}_1^{\gamma}\,,
\end{equation}
that are well defined on the domain of $Q_1^{\gamma}$
(we remind that $\gamma$ acquires one of the two values (\ref{gamma})).
They are nonlocal due to the presence of the operator of
rotations
\begin{equation}\label{RdefJ}
    R=e^{i\pi (J-\alpha)},\qquad R\varphi R=\varphi+\pi,
    \qquad R^2=1\,.
\end{equation}

The operators (\ref{hiddenSUSY})
anti-commute with $R$ while the Hamiltonian commutes with
this operator. The operators $\tilde{Q}_1$ and
$\tilde{Q}_2$  satisfy the relation
 \begin{equation}\label{hiddensusy}
      \{\tilde{Q}_a^{\gamma},
      \tilde{Q}_b^{\gamma}\}=2\delta_{ab}
      \mathcal{H}_{\alpha}^{\gamma}\,,
 \end{equation}
and form an alternative representation of the $N=2$
supersymmetry. As the supercharges (\ref{hiddenSUSY})
originate from the hidden supersymmetry of the spinless AB
systems, we call the operators (\ref{hiddenSUSY}) the
supercharges of the \textit{hidden supersymmetry}
(\ref{hiddensusy}).

Let us briefly discuss the action of the
supercharge $\tilde{Q}_1^\gamma$ on the eigenstates of
$\mathcal{H}^{0}_\alpha$ ($\mathcal{H}^{\pi}_\alpha$). It
is instructive to rewrite the diagonal component of
(\ref{hiddenSUSY}) in the polar coordinates,
\begin{equation}\label{Qhidex}
     \mathcal{P}_1+ie^{i\gamma}R\mathcal{P}_2=
    -ie^{-i\varphi}\left(\partial_r+\frac{1}{r}J
    \right)\Pi_+^{\gamma}
    -ie^{i\varphi}\left(\partial_r-\frac{1}{r}J
    \right)
    \Pi_-^{\gamma}\,,\quad \Pi_\pm^{\gamma}=\frac{1}{2}
    (1\pm e^{i\gamma}R)\,,
\end{equation}
where the operators $\Pi_\pm^{\gamma}$, are projectors for
$\gamma=0\,,\pi$. A similar form for the component
$\mathcal{P}_1-ie^{i\gamma}R\mathcal{P}_2$ is obtained from
(\ref{Qhidex}) by the substitution $R\rightarrow -R$. This
suggests that both $\tilde{Q}_1^\gamma$ and
$\tilde{Q}_2^\gamma$ preserve the subspaces
$\tilde{\mathfrak{H}}_{kl}$ of the form
\begin{equation}\label{Qhidsub}
    \tilde{\mathfrak{H}}_{kl}=\left\{f\,\vert\, f\in\left(\begin{array}{c}
    \{e^{i(2k-e^{i\gamma})\varphi},e^{i2k\varphi}\}\\
    \{e^{i(2l+e^{i\gamma})\varphi},e^{i2l\varphi}\}\end{array}\right)
    \otimes L_2(\mathbb{R}^+,rdr),\ k,\ l \in \mathbb{Z}\right\}.
\end{equation}
Then the action of the supercharge
$\tilde{Q}_1^{\gamma}$ on the wave functions
$\Psi_{E,l,\pm}$ can be inferred directly and reads
\begin{equation}\label{action2}
    \tilde{Q}_1^{\gamma}\Psi^{\gamma}_{E,2l,\pm}
    \sim\Psi^{\gamma}_{E,2l-1,\pm}\,,\qquad
    \tilde{Q}_1^{\gamma}\Psi_{E,2l-1,\pm}^{\gamma}
    \sim\Psi^{\gamma}_{E,2l,\pm}\,.
\end{equation}
Considering the states of zero energy, the situation is
qualitatively similar to the case of $Q_{1}^{\gamma}$. The Hamiltonian
$\mathcal{H}^{0}_\alpha$ ($\mathcal{H}^{\pi}_\alpha$) has
two spin-up and two spin-down zero modes in
$\tilde{\mathfrak{H}}_{kl}$ and $\tilde{Q}_1^\gamma$
annihilates only half of them.
\vskip0.05cm

We pose the following question\,: is it  possible to
include both the local,  $Q_1^{\gamma}$ and $Q_2^{\gamma}$,
and nonlocal, (\ref{hiddenSUSY}), supercharges in the
unifying scheme of an extended superalgebra? To respond it,
we note that all the operators  $Q_a^{\gamma}$ and
$\tilde{Q}_a^{\gamma}$ ($a=1,2$) have vanishing either
commutator or anti-commutator with each of the operators
\begin{equation}\label{3grading}
     \sigma_3\,,\qquad R\,,\qquad \sigma_3R\,.
\end{equation}
Each of the operators (\ref{3grading}) commutes with
$\mathcal{H}^{\gamma}_\alpha$.  Hence each is equally good
candidate for the grading operator $\Gamma$ of the extended
superalgebra, in which it should classify the integrals
into bosonic and fermionic ones.

Let us choose
\begin{equation}\label{Gamsig}
    \Gamma=\sigma_3\,,
\end{equation}
and treat this case in some detail;  on other  choices of
$\Gamma$ we will comment below. Besides
$\mathcal{H}_{\alpha}^{\gamma}$, $J$ and (\ref{3grading}),
we have other four nontrivial  bosonic (commuting with
$\sigma_3$) operators
\begin{equation}\label{set1}
     \tilde{Q}_1^{\gamma}\,,\qquad \tilde{Q}^{\gamma}_2\,,
     \qquad \tilde{Q}_3^{\gamma}=i\sigma_3R\tilde{Q}_1^{\gamma}\,,
    \qquad \tilde{Q}_4^{\gamma}=\sigma_3\tilde{Q}_1^{\gamma}\,.
\end{equation}
The number of fermionic  (anti-commuting with $\sigma_3$)
operators can be extended similarly,
\begin{equation}\label{set2}
     Q_1^{\gamma}\,,\qquad
     Q^{\gamma}_2\,,\qquad
      Q_3^{\gamma}=iRQ_1^{\gamma}\,,\qquad
       Q_4^{\gamma}=R\sigma_3Q_1^{\gamma}\,.
\end{equation}
The nontrivial anti-commutation relations for (\ref{set2})
are
\begin{equation}\label{qqnontr1+}
    \{Q_{\underline{A}}^\gamma,Q_{\underline{A}}^\gamma\}=
    2\mathcal{H}_{\alpha}^{\gamma}\,,\quad
    \underline{A}=1,2,3,4,\qquad
    \{Q_{1}^\gamma,Q_{4}^\gamma\}=
    \{Q_{2}^\gamma,Q_{3}^\gamma\}=2\sigma_3 R
    \mathcal{H}_{\alpha}^{\gamma}\,.
\end{equation}
For the nontrivial commutation relations between the
integrals (\ref{set1}) we have
\begin{equation}\label{qqnontr2}
    [\tilde{Q}_{2}^\gamma,\tilde{Q}_{1}^\gamma]=
    [\tilde{Q}_{3}^\gamma,\tilde{Q}_{4}^\gamma]=2iR
    \mathcal{H}_{\alpha}^{\gamma}\,,\qquad
    [\tilde{Q}_{2}^\gamma,\tilde{Q}_{4}^\gamma]=
    [\tilde{Q}_{3}^\gamma,\tilde{Q}_{1}^\gamma]=2i\sigma_3R
    \mathcal{H}_{\alpha}^{\gamma}\,.
\end{equation}
As soon as we require the superalgebra to be closed, we
have to calculate also the commutators between bosonic,
(\ref{set1}),  and fermionic,   (\ref{set2}), operators. In
this way, we get nonzero commutation relations
\begin{equation}\label{qtildq}
    [\tilde{Q}_1^\gamma,Q_3^\gamma]=-[\tilde{Q}_2^\gamma,Q_1^\gamma]
    =-2i\mathcal{W}_4^{\gamma}\,,\qquad
    [\tilde{Q}_1^\gamma,Q_4^\gamma]=[\tilde{Q}_2^\gamma,Q_2^\gamma]
    =2i\mathcal{W}_2^{\gamma}\,,\quad
\end{equation}
\begin{equation}\label{qtildq}
    [\tilde{Q}_3^\gamma,Q_3^\gamma]=[\tilde{Q}_4^\gamma,Q_1^\gamma]
    =-2i\mathcal{W}_3^{\gamma}\,,\qquad
    [\tilde{Q}_3^\gamma,Q_4^\gamma]=-[\tilde{Q}_4^\gamma,Q_2^\gamma]
    =-2i\mathcal{W}_1^{\gamma}\,,
\end{equation}
where
\begin{equation}\label{set3}
     \mathcal{W}_1^{\gamma}=\frac{i}{2}[\tilde{Q}_3^{\gamma}\,,
     {Q}_4^{\gamma}]=Q_1^{\gamma}\, \tilde{Q}_1^{\gamma}\,,\qquad
    \mathcal{W}^{\gamma}_2=iR\sigma_3\mathcal{W}_1^{\gamma}\,,\qquad
    \mathcal{W}^{\gamma}_3
    =i\sigma_3\mathcal{W}_1^{\gamma}\,,\qquad
    \mathcal{W}^{\gamma}_4=R\mathcal{W}_1^{\gamma}\,.
\end{equation}
Operators (\ref{set3}) anti-commute with (\ref{Gamsig}),
and have to be treated as a new set of independent
fermionic operators. The nontrivial anti-commutators
between them are
\begin{equation}\label{WWnontr}
    \{\mathcal{W}^{\gamma}_{\underline{A}},
    \mathcal{W}^{\gamma}_{\underline{A}}\}=
    2(\mathcal{H}_{\alpha}^{\gamma})^2\,,\quad
    \underline{A}=1,2,3,4,\qquad
    \{\mathcal{W}^{\gamma}_{1},
    \mathcal{W}^{\gamma}_{4}\}=
    \{\mathcal{W}^{\gamma}_{2},
    \mathcal{W}^{\gamma}_{3}\}=2R
    (\mathcal{H}_{\alpha}^{\gamma})^2\,.
\end{equation}
{}From here it follows, particularly, that the pair of the
integrals $\mathcal{W}^{\gamma}_{1}$ and
$\mathcal{W}^{\gamma}_{2}$ [as well as each of the pairs
($\mathcal{W}^{\gamma}_{3}$, $\mathcal{W}^{\gamma}_{4}$),
($\mathcal{W}^{\gamma}_{1}$, $\mathcal{W}^{\gamma}_{3}$)
and ($\mathcal{W}^{\gamma}_{2}$,
$\mathcal{W}^{\gamma}_{4}$)] generates the relations of the
$N=2$ nonlinear (second order) supersymmetry
\cite{hidsusynon}, \cite{NonSUSY},
\begin{equation}\label{nonlinsusy}
    \{\mathcal{W}^{\gamma}_a,\mathcal{W}^{\gamma}_b\}=
    2\delta_{ab}(\mathcal{H}_{\alpha}^{\gamma})^2\,,
    \quad a,b=1,2\,,\qquad
    [\mathcal{W}^{\gamma}_a,\mathcal{H}_{\alpha}^{\gamma}]=0\,.
\end{equation}

In the (anti)-commutators of (\ref{set3}) with either
(\ref{set1}) or (\ref{set2}), we get the products of
$\mathcal{H}_{\alpha}^{\gamma}$ with either (\ref{set2}) or
(\ref{set1}), e.g.,
\begin{equation}
     \{Q_3^{\gamma},\mathcal{W}^{\gamma}_1\}=
     -\{Q_4^{\gamma},\mathcal{W}^{\gamma}_3\}=2
     \tilde{Q}_2^{\gamma}\mathcal{H}_{\alpha}^{\gamma}\,,\qquad
     [\tilde{Q}_1^{\gamma},\mathcal{W}^{\gamma}_4]=
     [\tilde{Q}_3^{\gamma},\mathcal{W}^{\gamma}_3]=2i
     Q_3^{\gamma}\mathcal{H}_{\alpha}^{\gamma}\,.
\end{equation}
The missing commutation relations of the operators
(\ref{set1}), (\ref{set2}) and (\ref{set3}) with the
integral $J$ can easily be calculated by noticing that all
these operators commute with
\begin{equation}\label{centerZ}
     \mathcal{Z}^{\gamma}=J+
     \sigma_3\Pi_-^{\gamma}\,,
     \qquad [\mathcal{Z}^{\gamma},Q_{\underline{A}}^{\gamma}]=
     [\mathcal{Z}^{\gamma},\tilde{Q}_{\underline{A}}^{\gamma}]=
     [\mathcal{Z}^{\gamma},{W}_{\underline{A}}^{\gamma}]=0,
     \quad \underline{A}=1,2,3,4\,.
 \end{equation}
The complete superalgebra is, therefore, nonlinear due to
the presence of the Hamiltonian and operators $R$ and
$\sigma_3 R$ on the right hand side of some
(anti)-commutation relations, that is similar to the
nonlinearity of the symmetry algebra appearing in the
quantum Kepler problem and associated there with a hidden
symmetry provided by the Laplace-Runge-Lenz vector
\cite{Pauli}. The operator $\mathcal{Z}^{\gamma}$ plays
here the role of the central charge.

For other choices of the grading operator, $\Gamma=R$ or
$\Gamma=\sigma_3 R$, the sets of the supercharges
(\ref{set1}), (\ref{set2}) and (\ref{set3}) permute in the
role of the bosonic operators (one can check that the
integrals (\ref{set2}) commute with $\sigma_3R$ and those
from (\ref{set3}) commute with $R$). In the table, we
illustrate the separation of the operators into fermionic
and bosonic families.

\vspace*{2mm}
\begin{table}[ht]
\begin{center}\renewcommand{\arraystretch}{1.6}
\begin{tabular}{| c | c| c | }\hline
  Grading operator & Bosonic operators& Fermionic operators
  \\\hline\hline\\[-22pt]
  $\sigma_3$&$\mathcal{H}^{\gamma}_\alpha,\
  \mathcal{Z}^{\gamma},\ \sigma_3,\ R,\ \sigma_3R$,
  $\tilde{Q}_{\underline{A}}^{\gamma}$ & $
  Q_{\underline{A}}^{\gamma}$,
  $\mathcal{W}_{\underline{A}}^{\gamma}$ \\[3pt]
  \hline\\[-22pt]
  $R$&$\mathcal{H}^{\gamma}_\alpha,\ \mathcal{Z}^{\gamma},\
  \sigma_3,\ R,\ \sigma_3R$,
  $\mathcal{W}_{\underline{A}}^{\gamma}$ &
  $\tilde{Q}_{\underline{A}}^{\gamma}$,
  $Q_{\underline{A}}^{\gamma}$ \\[3pt]
  \hline
  $\sigma_3R$&$\mathcal{H}^{\gamma}_\alpha,
  \ \mathcal{Z}^{\gamma},\ \sigma_3,\ R,\ \sigma_3R$,
  $Q_{\underline{A}}^{\gamma}$
  & $\tilde{Q}_{\underline{A}}^{\gamma}$,
  $\mathcal{W}_{\underline{A}}^{\gamma}$ \\[3pt]
  \hline

\end{tabular}
\end{center}
\end{table}

\noindent The superalgebra remains qualitatively the same
for any choice of $\Gamma$. The operators $R$ and
$\sigma_3R$ that appear in some (anti-)commutation
relations for the case (\ref{Gamsig}), are changed for
$\sigma_3$ and $\sigma_3 R$ in the case $\Gamma=R$, and for
$\sigma_3$ and $R$ for $\Gamma=\sigma_3 R$. Such a
superalgebraic structure, characterized by the three
possible choices of the grading operators and three sets of
the supercharges, was  observed in the finite-gap periodic
quantum systems \cite{finitegap}, and was named
\textit{tri-supersymmetry}, see also \cite{DiracDelta}.

\begin{figure}[h!]
\begin{center}
\includegraphics[width=0.5\linewidth]{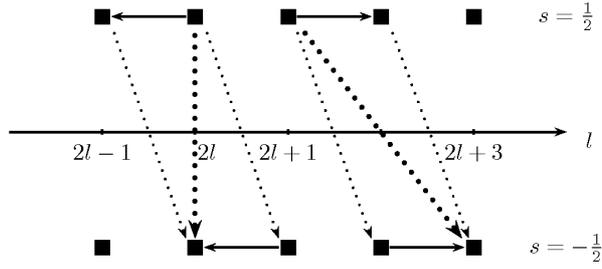}
\caption{The action of the operator $\mathcal{W}_1^0$
 (thick dotted arrows) on the states $\Psi_{E,2l,+}$ and
 $\Psi_{E,2l+1,+}$ as a sequential action of $\tilde{Q}_1^0$
 (solid arrows) and $Q_1^0$ (thin dotted arrows).
 Black squares represent the eigenstates $\Psi_{E,l,s}$
 with corresponding values of $l$ and $s$.}\label{fig1}
\end{center}
\end{figure}

The action of the supercharge $\mathcal{W}_1^{\gamma}$ on
the eigenvectors $\Psi_{E,l,\pm}$ can be deduced directly
from the relations (\ref{action1}) and (\ref{action2}),
keeping in mind the definition of
$\mathcal{W}^{\gamma}_{1}$ and
$[Q_1^{\gamma},\tilde{Q}_1^{\gamma}]=0$. We recall that the
operator $Q_1^{\gamma}$ preserves the subspaces
$\mathfrak{H}_{k}$ while the supercharge
$\tilde{Q}^{\gamma}_1$ preserves the subspaces
$\tilde{\mathfrak{H}}_{kl}$; the explicit action depends on
the values of the quantum numbers $l$ and $s$ in
$\Psi_{E,l,s}$. Let us write down as an example
\begin{equation}\label{example}
    \mathcal{W}_1^{\gamma}\Psi_{E,2l,+}^{\gamma}=Q_1^{\gamma}\tilde{Q}_1^{\gamma}
    \Psi_{E,2l,+}^{\gamma}=\tilde{Q}_1^{\gamma}Q_1^{\gamma}
    \Psi_{E,2l,+}^{\gamma}\sim\Psi_{E,2l,-}^{\gamma}\,,
    \end{equation}
which is illustrated schematically in Fig.\ref{fig1}

Notice that for the choice (\ref{Gamsig}) of the grading
operator, each of the four pairs of the odd supercharges
(\ref{set2})
\begin{equation}\label{QpairsH}
    (Q_1^{\gamma},Q_2^{\gamma}),\qquad
    (Q_3^{\gamma},Q_4^{\gamma}),\qquad
    (Q_1^{\gamma},Q_3^{\gamma}),\qquad
    (Q_2^{\gamma},Q_4^{\gamma})
\end{equation}
generates the $N=2$ supersymmetry, see Eq.
(\ref{qqnontr1+}). Similarly, the two pairs
\begin{equation}\label{QpairsHdef}
    (Q_{1+}^{\gamma},Q_{1-}^{\gamma}),\qquad
    (Q_{2+}^{\gamma},Q_{2-}^{\gamma})
\end{equation}
generate the deformed $N=2$ supersymmetry of the form
\begin{equation}\label{Hdefpi}
    \{Q_{as}^{\gamma},Q_{as'}^{\gamma}\}=
    2\delta_{ss'}\hat{\Pi}_s\mathcal{H}_\alpha^\gamma,\qquad
    a=1,2,\quad
    s,s'=+,-\,.
\end{equation}
Here $Q_{1\pm}^{\gamma}=\frac{1}{2}(Q_{1}^{\gamma}\pm
Q_{4}^{\gamma})=\hat{\Pi}_\pm Q_{1}^{\gamma}$,
$Q_{2\pm}^{\gamma}=\frac{1}{2}(Q_{2}^{\gamma}\pm
Q_{3}^{\gamma})=\hat{\Pi}_\pm Q_{2}^{\gamma}$, and
$\hat{\Pi}_\pm=\frac{1}{2}(1\pm \sigma_3 R)$ are the
projector operators. Analogously, as we noted above, the
integrals (\ref{set3}) generate the nonlinear [deformed for
the pairs $\mathcal{W}_{a\pm}^{\gamma}$, $a=1,2$,
constructed similarly to $Q_{a\pm}^{\gamma}$]
 $N=2$ supersymmetry. In this case, the projector operators
$\hat{\Pi}_\pm$ are changed for the projectors
$\Pi_\pm=\frac{1}{2}(1\pm R)$.

A similar picture  is valid for the corresponding fermionic
supercharges when we choose $\Gamma=R$ or
$\Gamma=\sigma_3R$.\vskip0.05cm

In summary of this Section, the spin$-1/2$ Aharonov-Bohm
system possesses $N=2$ supersymmetry in two cases only,
represented by $\mathcal{H}^{0}_{\alpha}$ and
$\mathcal{H}_{\alpha}^{\pi}$. It comes hand-in-hand with
the hidden supersymmetry represented by the supercharges
(\ref{hiddenSUSY}). Both, the standard and the hidden
supersymmetries, are unified in the framework of the
tri-supersymmetry.

This structure will reveal itself later on in the existence
of the three different types of the super-extended anyon
systems.

\section{Tri-supersymmetry and superconformal symmetry}\label{dynamical}

Besides the usual symmetries, the system  possesses
dynamical symmetries as well. The Hamiltonian
$\mathcal{H}_{\alpha}^{\gamma}$ together with the
generators of special conformal transformations
(expansions)~\footnote{To simplify notations, we omit
indexes $\alpha$ and $\gamma$ in dynamical integrals $D$
and $K$.}, $K$, and dilatations, $D$,
\begin{equation}\label{KD}
     K=\vec{X}{}^2
     \,,
     \qquad D=-\frac{1}{2}
     \left(\vec{X}\vec{\cal{P}}+\vec{\cal{P}}\vec{X}\right)\,,
     \quad \mbox{where}\quad X_j=x_j-2t{\cal P}_j,\quad
     j=1,2\,,
\end{equation}
form the  $so(2,1)$ Lie algebra
\begin{equation}\label{so(2,1)}
     [D,K]=2iK\,,\qquad
     [\mathcal{H}_{\alpha}^{\gamma},K]=4iD\,,
     \qquad [\mathcal{H}_{\alpha}^{\gamma},D]=
     2i\mathcal{H}_{\alpha}^{\gamma}\,.
\end{equation}
The operators (\ref{KD}) satisfy the relation
$\frac{d}{dt}\mathcal{O}=\partial_t\mathcal{O}-i
[\mathcal{H}_{\alpha}^{\gamma},\mathcal{O}]=0$
($\mathcal{O}$ being either $K$ or $D$), that justifies to
call them \textit{dynamical} symmetries.
Let us note that the Lie algebra $so(2,1)$ of dynamical symmetries in the
context of (spinless) Aharonov-Bohm system was first observed in
\cite{Jackiw}.

As long as $\gamma=0$ or $\pi$, neither $K$ nor $D$ alter
the asymptotic behavior of the wave functions prescribed by
(\ref{boundary}). Hence, both  operators preserve the
domain of $\mathcal{H}^{\gamma}_{\alpha}$. In particular,
invariance of the domain of $\mathcal{H}_{\alpha}^{\gamma}$
under the action of $D$ is associated with \textit{ the
scale invariance} of the  system. The relations
(\ref{so(2,1)}) establish then the conformal symmetry of
the model.\vskip0.05cm

Now, we discuss how the dynamical symmetries
(\ref{so(2,1)}) can be incorporated into the framework of
the tri-supersymmetry. The operators $K$ and $D$ are even
with respect to any of the possible grading operators
(\ref{3grading}). First, we consider the case
$\Gamma=\sigma_3$, and discuss the structure of the
superalgebra that includes $K$ and $D$.

The odd supercharges of tri-supersymmetry are indentified
with (\ref{set2}) and (\ref{set3}), see the table. Let us
focus on (\ref{set2}). It is well known \cite{Park},
\cite{Horvathy} that the usual $N=2$ supersymmetry
associated with the supercharges $Q_1^{\gamma}$ and
$Q_{2}^{\gamma}$ can be expanded into the superalgebra
$osp(2|2)$. The same, \emph{up to the deformations},
algebraic structure can be obtained when we expand by $K$
and $D$ the $N=2$ supersymmetry associated with any other
pair of the supercharges from (\ref{QpairsH}),
(\ref{QpairsHdef}). The (anti-)commutation relations of the
(deformed) superconformal algebra give rise to the
additional dynamical symmetries,
\begin{equation}\label{B}
      [Q_{\underline{A}}^{\gamma},K]=
      -2iS_{\underline{A}}^{\gamma}\,,\quad
      {\rm where}\quad
      S_1^\gamma=X_1\sigma_1+X_2\sigma_2\,,
\end{equation}
and $S_2^\gamma$, $S_3^\gamma$ and $S_3^\gamma$ are related
to $S_1^\gamma$ similarly to (\ref{set2}). The
$S_{\underline{A}}^\gamma$ satisfy the anti-commutation
relations
\begin{equation}\label{qqnontr1}
    \{S_{\underline{A}}^\gamma,S_{\underline{A}}^\gamma\}=
    2K\,,\quad
    \underline{A}=1,2,3,4,\qquad
    \{S_{1}^\gamma,S_{4}^\gamma\}=
    \{S_{2}^\gamma,S_{3}^\gamma\}=2\sigma_3 R
    K\,.
\end{equation}
 For the four odd integrals
$Q_{\underline{A}}^\gamma$ and $S_{\underline{A}}^\gamma$
with $\underline{A}=a=1,2$, we have the nontrivial
anti-commutation relations in addition to those presented
above,
\begin{equation}\label{QSanti}
    \{S_{a}^\gamma,Q_{b}^\gamma\}=-
    2\delta_{ab}D+2\epsilon_{ab}{\cal J}\,,
\end{equation}
\begin{equation}\label{HSg}
    [\mathcal{H}_\alpha^\gamma,S_{a}^\gamma]=
    -2iQ_{a}^\gamma\,,\quad
    [K,Q_{a}^\gamma]=2iS_{a}^\gamma\,,\quad
    [D,Q_{a}^\gamma]=-iQ_{a}^\gamma\,,\quad
    [D,S_{a}^\gamma]=iS_{a}^\gamma\,,
\end{equation}
\begin{equation}\label{JsQS}
    [{\cal J},Q_{a}^\gamma]=-i\epsilon_{ab}Q_{b}^\gamma \,,\qquad
    [{\cal J},S_{a}^\gamma]=-i\epsilon_{ab}S_{b}^\gamma\,,
\end{equation}
where ${\cal J}=J+\sigma_3$, which correspond to the
$osp(2|2)$ superalgebra.

In the case of the odd generators
$Q_{\underline{A}}^\gamma$ and $S_{\underline{A}}^\gamma$
with $\underline{A}=3,4$ we have the same $osp(2|2)$
superalgebra with the change of the indices $1\rightarrow
4$,  $2\rightarrow 3$ in correspondence with relations
$Q_{\underline{4}}^\gamma=\sigma_3RQ_{\underline{1}}^\gamma$,
$Q_{\underline{3}}^\gamma=\sigma_3RQ_{\underline{2}}^\gamma$,
and the same relations for the integrals $S^\gamma$.

The other two pairs of $Q_{\underline{A}}^\gamma$ from
(\ref{QpairsH}), and the two pairs in (\ref{QpairsHdef})
can be extended in the similar vain, giving rise to
corresponding pairs of $S_{\underline{A}}^\gamma$, and the
pairs $(S_{a+}^\gamma$,  $S^\gamma_{a-}$), $a=1,2$,
respectively. The resulting algebras are of the same form
as (\ref{QSanti})--(\ref{JsQS}), but with some of the
(anti)-commutation relations to be deformed by inclusion of
the factor $\sigma_3R$, or the projector as in Eq.
(\ref{Hdefpi}). For instance,
$\{S_1^\gamma,Q_3^\gamma\}=2\sigma_3R\,{\cal J}$, cf.
(\ref{QSanti}). Hence, we get six different (deformed)
representations of the superalgebra $osp(2|2)$ that are
included in the finite-dimensional extension of the
superalgebra $osp(2|2)$ generated by
$\mathcal{H}^{\gamma}_{\alpha}$, $K$, $D$ and the
supercharges $Q^{\gamma}_{\underline{A}}$. Note that the
operator $\sigma_3R$ commutes with all the generators of
the $osp(2|2)$ here.\vskip0.05cm

Similarly to (\ref{B}), the commutators of (\ref{set1})
with $K$ generate a new set of dynamical integrals,
\begin{equation}\label{Btil}
      [\tilde{Q}_{\underline{A}}^{\gamma},K]=
      -2i\tilde{S}_{\underline{A}}^{\gamma}\,,
\end{equation}
which are fermionic operators for $\Gamma=\sigma_3$. But
when we expand here the set of the even,
$\mathcal{H}_\alpha^\gamma$, $K$, $D$, ${\cal J}$, and odd,
$Q_{\underline{A}}^\gamma$, $S_{\underline{A}}^\gamma$,
generators with the bosonic  operators
$\tilde{Q}_{\underline{A}}^\gamma$ or/and fermionic
operators $\mathcal{W}_{\underline{A}}^\gamma$, an infinite
number of the new operators appears.\vskip0.05cm

The picture is similar for other choices of the grading
operator. For instance, for $\Gamma=R$,  the (deformed)
finite-dimensional extension of the $osp(2|2)$ superalgebra
is obtained if we supply the generators of the conformal
symmetry with one (not both) of the two sets of the
fermionic operators, $Q_{\underline{A}}^{\gamma}$ or
$\tilde{Q}_{\underline{A}}^{\gamma}$. The same is valid for
the case $\Gamma=\sigma_3R$, when the $so(2,1)$ generators
are supplied by the fermionic operators
$\tilde{Q}_{\underline{A}}^{\gamma}$. As soon, however, as
we supply the $so(2,1)$ generators (\ref{KD}) with the
integrals $\mathcal{W}_{\underline{A}}^\gamma$, we get the
infinite superalgebra.\vskip0.1cm

Hence, the extension of the tri-supersymmetry by the
dynamical integrals $D$ and $K$  gives rise to the infinite
superalgebra for any choice of the grading operator. As we
saw, six (deformed) superalgebras $osp(2|2)$ are included
as finite subalgebras. Note that the infinite superalgebra
still has the central element $\mathcal{Z}^{\gamma}$, see
(\ref{centerZ}).

\section{Tri-supersymmetry and three types of supersymmetric anyons}

We provide now an alternative interpretation of the results
presented in the previous Sections, applying them to the
theory of anyons. We elaborate the idea for the system
described by $\mathcal{H}_{\alpha}^{0}$. The case of
$\mathcal{H}_{\alpha}^{\pi}$ can be treated  similarly.

The dynamical realization of the anyons \cite{LeiMyr} that
was proposed by Wilczek  in \cite{wilczek}, is based, in
fact, on the Aharonov-Bohm effect. In such a picture, anyon
is considered as a ``composite'', statistically charged
particle that is either boson or fermion, to which a
magnetic vortex is attached. The presence of the vortex
provides the peculiar statistical properties of the anyons.
Consider the system of \emph{two identical non-relativistic
anyons} in such a picture. It is assumed that each particle
feels only the potential produced by the vortex attached to
the other particle. The Hamiltonian of the system we denote
as
\begin{equation}\label{Hany}
    H_{any}=
   2\sum_{I=1}^2\left(\vec{p}_I-\vec{a}_I(\vec{r})
    \right)^2\,,
\end{equation}
see \cite{anyonsbooks}. Index $I\,$ labels the individual
particles (whose masses are $m_1=m_2=4$) with the momenta
$\vec{p}_I=-i\partial/\partial \vec{x}_I$. The vector
$\vec{r}\,$ is a relative coordinate of the particles,
$\vec{r}=\vec{x}_1-\vec{x}_2$. The potentials $\vec{a}_I$,
\begin{equation}\label{anyA}
    a_1^k(\vec{r})= - a_2^k(\vec{r})=
    \frac{1}{2}\alpha
    \epsilon^{kl}\frac{r^l}{\vec{r}\,{}{}^2}\,,
\end{equation}
encode the ``statistical'' interaction of the particles.
When we write the Hamiltonian in the center-of-the-mass
coordinates, the relative motion of the particles is
governed by the effective Hamiltonian
\begin{equation}\label{Hanyrel}
    H_{rel}=-\partial_{r}^2-\frac{1}{r}
    \partial_{r}+\frac{1}{r^2}(-i\partial_{\varphi}+\alpha)^2\,,
\end{equation}
 where
$r$ and $\varphi$ are the polar coordinates of the
$\vec{r}$.

Formally, this operator coincides with the spinless
Hamiltonians $H_{\alpha}^0$ and $H_{\alpha}^{AB}$ of
(\ref{h0}). But its domain is quite different. The domain
of (\ref{Hanyrel}) reflects the nature of the anyons and is
composed of functions which are either symmetric or
anti-symmetric under the change $\varphi\rightarrow
\varphi+\pi$. The periodicity of wave functions depends, in
turn, on the nature of the interacting particles. The
two-body wave function has to be symmetric in $\pi$
(invariant with respect to the exchange of the particles)
as long as we deal with anyons based on bosons. When the
vortices are attached to fermions, the wave function has to
change its sign after the substitution $\varphi\rightarrow
\varphi+\pi$. Hence, we have
\begin{equation}\label{nature}
     \psi_{\alpha}(r,\varphi)=\sum_le^{il\varphi}f_{\alpha,l}(r),
     \quad l\in\left\{\begin{array}{l}2\mathbb{Z}\
     \mbox{for anyons based on bosons\,,}\\
    2\mathbb{Z}+1\ \mbox{for anyons based on fermions\,.}
    \end{array}\right.
\end{equation}
Notice that the system can be described alternatively by
the free Hamiltonian
$H_{rel}=-\partial_{r}^2-1/r\partial_r+(-i\partial_{\varphi})^2/r^2$.
However, the wave functions  have to acquire the gauge
factor $e^{i\alpha\varphi}$ in this case to keep the
description equivalent to (\ref{Hanyrel}) and
(\ref{nature}), i.e. $\psi_{\alpha}(r,\varphi)=
e^{i\varphi\alpha}\sum_le^{il\varphi}f_{\alpha,l}(r)$. In
this alternative picture, after the substitution
$\varphi\rightarrow \varphi+\pi$ the wave functions acquire
the phase $e^{i\pi\alpha}$  which interpolates between the
values corresponding to Bose and Fermi statistics. We
prefer to use the framework (\ref{nature}), where the wave
functions are $2\pi$-periodic for any value of $\alpha$.
\vskip0.1cm

Let us return to our current system. Instead  to
treat its Hamiltonian as that of Pauli, we consider it as a
direct sum of four operators, just as the $4\times 4$
matrix operator
\begin{equation}\label{hany}
     \mathcal{H}^{\gamma=0}_{\alpha}=diag\,(
    H^{0}_{\alpha,+},H^{0}_{\alpha,-},H^{AB}_{
     \alpha,+},H^{AB}_{\alpha,-})\,,
     \qquad H^{0}_{\alpha,\pm}=H^{0}_{\alpha}\Pi_{\pm},
     \quad H^{AB}_{\alpha,\pm}=H^{AB}_{\alpha}\Pi_{\pm}\,,
\end{equation}
where we used the projectors $\Pi_{\pm}=\frac{1}{2}(1\pm
R)$ which separate symmetric and anti-symmetric under
displacement of $\varphi$ in $\pi$ wave functions. 
We remind that the nonzero elements of this diagonal
Hamiltonian operator coincide as differential operators but
can differ in their domains. The singular wave functions
are contained in the domain of $H^{0}_{\alpha,-}$ only. The
operator $H^{AB}_{\alpha,-}$ acts on regular functions
which are anti-periodic in $\pi$. The operators
$H^{0}_{\alpha,+}$ and $H^{AB}_{\alpha,+}$ coincide
actually in their domains and describe the same physical
setting.\vskip0.1cm

Now, each of the operators $H^{AB}_{\alpha,\pm}$ and
$H^0_{\alpha,\pm}$ can be interpreted as the Hamiltonian of
the relative motion of the two identical anyons. Indeed,
they coincide formally with (\ref{Hanyrel}) and their
domains are composed of either odd or even partial waves
(\ref{nature}). The regularity of the wave functions of
$H_{\alpha,\pm}^{AB}$ and $H_{\alpha,+}^0$ can be
understood as a manifestation of the hard core interaction
of the anyons. The singular behavior at $r\sim 0$ in the
domain of $H_{\alpha,-}^0$ can be reinterpreted as the
nontrivial contact interaction of the particles
\cite{manueltarrach}. Notice that this interaction appears
for one partial wave only, all other partial waves are
regular at the origin.

We conclude, therefore, that $H^{AB}_{\alpha,+}$ and
$H^0_{\alpha,+}$ describe two anyons based on bosons.
Similarly, $H_{\alpha,-}^{AB}$ and $H^0_{\alpha,-}$
describe the systems of two anyons based on fermions.

We identify now the sense of the cornerstones of the
tri-supersymmetry, the integrals of motion  $Q_{1}^0$,
$\tilde{Q}_{1}^0$ and $\mathcal{W}_1^{0}$, in this
framework. Keeping in mind (\ref{4}) and (\ref{hiddenSUSY})
and the splitting of $\mathcal{D}(\mathcal{H}_{\alpha}^0)$
which was made implicitly in (\ref{hany}), we can write
\begin{equation}\label{Qanyon}
     Q_1^{0}=\left(\begin{array}{cccc}0&0&0&q_+\\0&0&q_+&
     0\\0&q_-&0&0\\q_-&0&0&0\end{array}\right),\,\,
     \tilde{Q}_1^0=\left(\begin{array}{cccc}0&q_-&0&0\\q_+
     &0&0&0\\0&0&0&q_+\\0&0&q_-&0\end{array}\right),\,\,
     \mathcal{W}_1^0=\left(\begin{array}{cccc}0&0&q_+q_-&0
     \\0&0&0&q_+^2\\q_-q_+&0&0&0\\0&q_-^2&0&0\end{array}\right),
\end{equation}
where $q_\pm$ are defined in (\ref{4}). The nontrivial
matrix elements of the commutation relation
$[Q_1^0,\mathcal{H}^{0}_{\alpha}]=0$ give
\begin{equation}\label{int1}
    H_{\pm}^0q_{+}=q_{+}H_{\mp}^{AB}\,,\qquad
    H_{\pm}^{AB}q_{-}=q_{-}H_{\mp}^{0}\,,
\end{equation}
where $H_{\pm}^0=H_{\alpha,\pm}^0$,
$H_{\pm}^{AB}=H_{\alpha,\pm}^{AB}$.
 The commutator $[\tilde{Q}_1^0,
 \mathcal{H}^{0}_{\alpha}]=0$ gives rise to the operator equations
\begin{equation}\label{int2}
     H_{\pm}^0q_{\mp}=q_{\mp}H_{\mp}^{0}\,,
     \qquad H_{\pm}^{AB}q_{\pm}=q_{\pm}H_{\mp}^{AB}\,,
\end{equation}
and the commutator $[\mathcal{W}_1^0,
\mathcal{H}_{\alpha}^0]=0$ can be rewritten as
\begin{equation}\label{int3}
    H_{-}^{AB}q_{-}^2=q_{-}^2H_{-}^{0},\qquad
    H_{-}^0q_{+}^2=q_{+}^2H_{-}^{AB},\quad
    H_{+}^{AB}q_{-}q_+=q_{-}q_+H_{+}^{0},\quad
    H_{+}^0q_{+}q_-=q_{+}q_-H_{+}^{AB}.
\end{equation}
Due to the identity $H_{\alpha,+}^0=H_{\alpha,+}^{AB}$, the
last  two relations tell that $q_+q_-=q_-q_+$ is a symmetry
of $H_{\alpha,+}^{AB}$ (as well as of $H_{\alpha,+}^0$).
However, it is trivial since $q_+q_-$ coincides with the
Hamiltonian. Not all the relations in
(\ref{int1})--(\ref{int3}) are independent. We reduce their
number using the identity
$H_{\alpha,+}^0=H_{\alpha,+}^{AB}$. The independent
relations are
\begin{equation}\label{susy1}
     H_+^0q_-=q_-H_{-}^0\,,\qquad q_+H_+^0=H_{-}^{0}q_+\,,
\end{equation}
\begin{equation}\label{susy2}
     H_+^0q_+=q_+H_{-}^{AB}\,,\qquad q_-H_+^0=H_{-}^{AB}q_-\,,
\end{equation}
\begin{equation}\label{susy3}
     H_-^{AB}q_-^2=q_-^2H_{-}^0\,,\qquad
     q_+^2H_-^{AB}=H_{-}^{0}q_-^2\,.
\end{equation}
Each of the relations (\ref{susy1}), (\ref{susy2})  can be
understood as  intertwining relation of the $N=2$ standard
supersymmetry \cite{CKS}. The relations (\ref{susy3}) give
rise to the nonlinear (the second order) supersymmetry
\cite{NonSUSY}. We have, therefore, three supersymmetric
systems, for which the superpartner Hamiltonians are
two-particle anyon systems.

The relations (\ref{susy1}) and (\ref{susy2})
give rise to the linear $N=2$ supersymmetry
\begin{equation}
     \left[\mathbf{q}^{(j)}_a,\mathbf{h}^{(j)}\right]=0,
     \quad\left\{\mathbf{q}^{(j)}_a,\mathbf{q}^{(j)}_b\right\}=
     2\delta_{a,b}\mathbf{h}^{(j)}\,,
     \quad j,a,b =1,2\,,
\end{equation}
represented, in case of (\ref{susy1}), by the
matrix Hamiltonians and corresponding supercharge operators
\begin{equation}\label{anysus1}
     \mathbf{h}^{(1)}=\left(\begin{array}{cc}H_{+}^0&0\\0&
     H_-^0\end{array}\right),\quad
    \mathbf{q}^{(1)}_1=
     \left(\begin{array}{cc}0&q_-\\q_+&0\end{array}\right),\quad
     \mathbf{q}^{(1)}_2=
     i\left(\begin{array}{cc}0&-q_-\\q_+&0\end{array}\right),
\end{equation}
while in the case of (\ref{susy2}) by the operators
\begin{equation}\label{anysus2}
     \mathbf{h}^{(2)}=\left(\begin{array}{cc}H_{+}^0&0\\0&H_-^{AB}
     \end{array}\right),\quad
     \mathbf{q}^{(2)}_1=\left(\begin{array}{cc}0&q_+\\q_-&0
     \end{array}\right),\quad
     \mathbf{q}^{(2)}_2=i\left(\begin{array}{cc}0&-q_+
     \\q_-&0\end{array}\right).
\end{equation}
The supercharges change the nature of the anyons in the
two-body systems; they transform the boson-based anyons of
$H_{+}^0$ into the fermion-based anyons of either $H_{-}^0$
or $H_{-}^{AB}$. In addition, they change the contact
interaction in $\mathbf{h}^{(1)}$ from the hard-core
interaction of $H_{+}^0$ to the nontrivial contact
interaction of $H_{-}^0$ (and vice versa).

The situation differs in the system associated with (\ref{susy3}).
The $N=2$ supersymmetry generated by the operators
\begin{equation}\label{anysus3}
     \mathbf{h}^{(3)}=\left(\begin{array}{cc}H_{-}^0&0\\0&H_-^{AB}
     \end{array}\right),\quad
     \mathbf{q}^{(3)}_2=\left(\begin{array}{cc}0&q_+^2\\q_-^2&0
     \end{array}\right),\quad
     \mathbf{q}^{(3)}_2=i\left(\begin{array}{cc}0&-
     q_+^2\\q_-^2&0\end{array}\right)
\end{equation}
 is nonlinear,
\begin{equation}
     \left[\mathbf{q}^{(3)}_{a},\mathbf{h}^{(3)}\right]=0,
\quad\left\{\mathbf{q}^{(3)}_{a},\mathbf{q}^{(3)}_b
     \right\}=2\delta_{ab}\left(\mathbf{h}^{(3)}
     \right)^2\,,\quad a,b=1,2\,.
\end{equation}
The supercharges preserve the nature of the anyons and just
alter the contact interaction of the two-body systems
described by $H_{-}^0$ and $H_{-}^{AB}$.

The role of the grading operator in all the three cases
(\ref{anysus1})--(\ref{anysus3}) is played by
$\sigma_3$.\vskip0.05cm

Concluding, coherently with the tri-supersymmetric
structure of the system $\mathcal{H}^0_\alpha$ described in
the previous Sections, the integrals $Q_1^0$,
$\tilde{Q}_1^0$ and $\mathcal{W}_1^0$ give rise to the
three different supersymmetric models
(\ref{anysus1})--(\ref{anysus3}) of the two-body systems of
interacting anyons. The first two models [each is composed
from the boson- and fermion-based anyons] are described by
a linear $N=2$ supersymmetric structure. The third one,
(\ref{anysus3}), composed from the two fermion-based anyon
subsystems, is described by the nonlinear, second order
$N=2$ supersymmetry. Due to the results of Section
\ref{dynamical}, the systems (\ref{anysus1}) and
(\ref{anysus2}) have the superconformal symmetry generated,
besides the supercharges and the Hamiltonian, by the
diagonal (even) operators $K$ and $D$, and by the odd
dynamical symmetries
$\mathbf{s}_1^{(j)}=i[\mathbf{q}^{(j)}_1,K]$,
$\mathbf{s}_2^{(j)}=i\sigma_3\mathbf{s}_1^{(j)}$. In
contrary, although the supersymmetric Hamiltonian
$\mathbf{h}^{(3)}$ has the conformal symmetry, its
supercharges cannot be included in the closed, finite Lie
superalgebra together with the dynamical symmetries $K$ and
$D$. As we saw in the Section \ref{dynamical}, the
superalgebra would be infinitely generated in this case.

The similar treatment applies to the system described by
$\mathcal{H}^{\pi}_{\alpha}$. In that case, the subsystems
$H_{-}^{AB}$ and $H_{-}^{\pi}$ would coincide.

\section{Discussion and outlook}

The system of the spin-1/2 particle in the field of the
magnetic vortex, that is described by the Hamiltonians
$\mathcal{H}_{\alpha}^{0}$ or $\mathcal{H}_{\alpha}^{\pi}$,
has a rich algebraic supersymmetry structure. We found that
the existence of the standard $N=2$ supersymmetry is
accompanied by the nonlocal supercharges (\ref{hiddenSUSY})
of the hidden supersymmetry. They form a different
realization of the $N=2$ supersymmetry of the model. Both
the local and nonlocal supercharges can be unified in the
framework of the tri-supersymmetry. There are three
possible candidates for the grading operator, see
(\ref{3grading}), and three sets (\ref{set1}), (\ref{set2})
and (\ref{set3}) of the operators which permute in the role
of the fermionic supercharges, dependently on the choice of
the grading operator.\vskip0.05cm

The tri-supersymmetry can be extended by the conformal
symmetry (\ref{so(2,1)}) of the model. The extension gives
rise to the infinitely generated superalgebra. It contains,
however, the (deformed) finite dimensional extension of the
superconformal symmetry $osp(2|2)$. \vskip0.05cm

We have applied the obtained results to the theory of
anyons, by reinterpreting the system and its algebraic
structure in terms of the supersymmetric two-body model of
the interacting anyons (\ref{anysus1})--(\ref{anysus3}).
Coherently with the described tri-supersymmetric structure,
the three different associated  anyon systems are
characterized by either linear, or \textit{non-linear}
$N=2$ supersymmetry graded by $\sigma_3$. \vskip0.05cm

The setting with integer magnetic flux, which is unitary
equivalent to a free particle case with $\alpha=0$, is
worth a separate note and a related comment on
translational invariance. As we observed, for $\alpha=0$
the system is specified uniquely, and its wave functions
(\ref{12}) and (\ref{14}) are regular at the origin. It has
the standard $N=2$ supersymmetry given by the supercharges
$Q_1$ and $Q_2$. The supercharges of the hidden
supersymmetry can be defined as well. The simplicity of the
$\alpha=0$ case, however, admits a greater freedom in their
definition; all the operators of the two-parameter family
\begin{equation}\label{Qfree}
    \tilde{Q}_{1,\epsilon_1,\epsilon_2}=diag(\mathcal{P}_1+\epsilon_1
    iR\mathcal{P}_2,\mathcal{P}_1+\epsilon_2
    iR\mathcal{P}_2),\quad
    \tilde{Q}_{2,\epsilon_1,\epsilon_2}=
    iR\tilde{Q}_{1,\epsilon_1,\epsilon_2},\quad
    \alpha=0,
\end{equation}
are well defined for $\epsilon_1, \epsilon_2=\pm 1$. The
set of integrals (\ref{Qfree}) is equivalent to the set
\begin{equation}
     diag(\mathcal{P}_j,0),\quad diag(iR\mathcal{P}_j,0),
     \quad diag(0,\mathcal{P}_j),\quad diag(0,iR\mathcal{P}_j),\quad j=1,2,
\end{equation}
that just manifests the translational invariance and the
reflection (rotation in $\pi$) symmetry of the spin-up and
spin-down components of the free-particle Hamiltonian. When
the magnetic flux is switched on, the translational
invariance of the system breaks down. Indeed, the
generators $\mathcal{P}_1$ and $\mathcal{P}_2$ are not
physical as they alter the boundary condition
(\ref{boundary}), and consequently their commutator with
$\mathcal{H}_{\alpha}^{\gamma}$ is not well defined. The
breakdown of translational invariance reduces the set of
symmetries (\ref{Qfree}) in half, leaving just the
operators which coincide with (\ref{hiddenSUSY}). In this
sense, the supercharges of the hidden supersymmetry
$\tilde{Q}_1^{\gamma}$ and $\tilde{Q}_2^{\gamma}$ can be
understood as the successors of the translational symmetry.
A deeper investigation of this point in the context of the
associated Galilei symmetry goes, however, beyond the scope
of the present paper and will be presented elsewhere.
\vskip0.05cm

In conclusion, it would be interesting to apply the
approach presented here to investigation of supersymmetry
in the system of the spin-1/2 particle in the field of the
magnetic monopole, where the issue of the domain of
definition is also essential \cite{Karat}, as well as in
the setting with several magnetic fluxes embedded into the
homogeneous magnetic field \cite{mine}, and to test them on
the presence of the hidden supersymmetry.\vskip0.3cm

\noindent \textbf{Acknowledgements.}
 The
work has been partially supported by DICYT (USACH), MECESUP
Project FSM0605, and FONDECYT Grants 1095027, 3100123 and
3085013, Chile, by CONICET (PIP 01787),  UNLP
(Proy.~11/X492) and ANPCyT (PICT-2007-00909), Argentina, by
the grant LC06002 of the Ministry of Education, Youth and
Sports of the Czech Republic, and by Spanish Ministerio de
Educaci\'on  under Project SAB2009-0181 (sabbatical grant
of MSP). H.F. is indebted to the Physics Department of
Santiago University (Chile) for hospitality. The Centro de
Estudios Cient\'{\i}ficos (CECS) is funded by the Chilean
Government through the Millennium Science Initiative and
the Centers of Excellence Base Financing Program of
Conicyt. CECS is also supported by a group of private
companies which at present includes Antofagasta Minerals,
Arauco, Empresas CMPC, Indura, Naviera Ultragas and
Telef\'onica del Sur.

\end{document}